\documentstyle[multicol,prl,aps]{revtex}
\begin{document}
\draft
\title{
A Viscoelastic Model of Phase Separation}
\author{Hajime Tanaka}
\address{
Institute of Industrial Science, University of Tokyo, Minato-ku, 
Tokyo 106, Japan. 
}
\date{Received 26 November 1996}
\maketitle
\begin{abstract}
{
We show here a general model of phase separation in isotropic condensed 
matter, namely, a {\it viscoelastic model}.  
We propose that the bulk mechanical 
relaxation modulus that has so far been ignored in previous theories 
plays an important role in viscoelastic phase separation 
in addition to the shear relaxation modulus.  
In polymer solutions, for example, attractive interactions between polymers 
under a poor-solvent condition likely cause the transient gellike behavior, 
which makes both bulk and shear modes active. 
Although such attractive interactions between molecules of the same component 
exist universally in the two-phase region of a mixture, 
the stress arising from attractive interactions is asymmetrically 
divided between the components only in dynamically asymmetric mixtures such as 
polymer solutions and colloidal suspensions. 
Thus, the interaction network between the slower components, 
which can store the elastic energy against its deformation 
through bulk and shear moduli, is formed. 
This unique feature originates from the difference in mobility 
between two components of a mixture. 
It is the bulk relaxation modulus associated with this interaction network 
that is primarily responsible for 
the appearance of the sponge structure peculiar to viscoelastic phase 
separation and the phase inversion: 
It suppresses the short-wavelength concentration fluctuations 
in the initial stage and causes the volume shrinking of a more viscoelastic 
phase. We also propose a simple general law 
of the stress division between the two components of a mixture, 
as a straightforward extension of that obtained in polymer mixtures.
We demonstrate that a viscoelastic model of phase separation 
including this new effect is a general model 
that can describe all types of isotropic phase separation 
including solid and fluid models as its special cases 
without any exception, if there is no coupling with 
additional order parameter. 
We show that this feature leads to a phenomenon of ``order-parameter 
switching'' during viscoelastic phase separation, even if it is driven 
by a single thermodynamic driving force.
The physical origin of volume shrinking behavior during viscoelastic phase 
separation and the universality of the resulting spongelike structure 
are also discussed. 
}

\end{abstract}

\begin{multicols}{2}

\section{Introduction}
Phase-separation phenomena are widely observed 
in various kinds of condensed matter including metals, semiconductors, 
simple liquids, and complex fluids such as polymers, surfactants, colloids, 
and biological materials. 
The phenomena play key roles in the pattern evolution 
of immiscible multi-component mixtures of any material. 
Thus, the phase-separation dynamics 
has been intensively studied in the past two decades 
from both the experimental 
and the theoretical viewpoints \cite{Gunton}.  
From the concept of dynamic universality, phase-separation 
phenomena have been classified into various theoretical models 
by Hohenberg and Halperin \cite{HH}: 
For example, phase separation in solids is known as ``solid model (model B)'', 
while phase separation in fluids as ``fluid model (model H)'' \cite{HH}. 
For the former the local concentration can be changed only by material 
diffusion, while for the latter by both diffusion and flow. 
It has been established that within each group the behavior is universal 
and does not depend on the details of material \cite{Gunton,HH}. 

In all conventional theories of critical phenomena and 
phase separation, however, 
the same dynamics for the two components of a binary mixture, 
which we call ``dynamic symmetry'' between the components, 
has been implicitly assumed \cite{Gunton,HH}. 
However, such an assumption of dynamic symmetry is hardly 
valid in various mixtures, especially in a material group 
of ``complex fluids''. 
Recently we have found that in mixtures  
having intrinsic "dynamic asymmetry" between its components 
({\it e.g.} polymer solutions composed of long chainlike molecules and simple liquid 
molecules), critical concentration fluctuation 
is not necessarily only the slow mode of the system 
and, thus, we have to consider the interplay between 
critical dynamics and the slow dynamics of material itself 
\cite{HT0,HT01,HT1,HT2}. 
In addition to a solid and a fluid model, thus, 
we need the third general model for 
phase separation in condensed matter, which we call a 
``viscoelastic model'' \cite{HT1,HT2}.  

To describe the difference in elementary dynamics between the 
two components of a mixture, we need a basic model 
that can treat the motion of each component separately: 
such a model is known as a "two-fluid model". 
The basic dynamic equations of viscoelastic model have been 
derived to understand the coupling between the stress and diffusion 
\cite{BD,TF,Helfand}
and also the unusual shear effects 
in polymer solutions \cite{Helfand,Onuki,Doi,DoiOnuki,Milner,OnukiN}, 
which are known as ``Reynolds effects'', on the basis 
of a two-fluid model \cite{BD,DG,Doi1}. 

In this paper, we propose that we need some essential modification 
to the ``viscoelastic model'' of phase separation described above: 
We believe \cite{HTG} that the bulk relaxation modulus, 
which has been neglected (or, more strictly, not treated as an important 
physical factor) 
in previous theories \cite{DoiOnuki,Milner}, plays an important role 
in viscoelastic phase separation as in gel phase separation. 
This modified viscoelastic model can describe any kinds of phase separation 
in mixtures of isotropic condensed matter without any exception, 
if there is no coupling with additional order parameter. 
In Sec. II, we review the theoretical derivation of a viscoelastic 
model based on a two-fluid model. In Sec. III, we discuss the 
internal modes of material itself and the resulting stress and 
how the stress is partitioned between the two components. In Sec. IV, 
we describe the basic equations of a viscoelastic model. In Sec. V, 
we discuss the origins of asymmetric stress division, using 
a few examples. We also propose a general rule of stress 
division that is independent of material. In Sec. VI, we discuss 
the generality of a viscoelastic model and demonstrate that 
all the models of phase separation in isotropic condensed matter 
are special cases of a viscoelastic model. 
In Sec. VII, we discuss the viscoelastic suppression of local 
concentration fluctuations in polymer solutions, focusing on 
the roles of bulk relaxation modulus. 
In Sec. VIII, we 
demonstrate that characteristic features of viscoelastic phase 
separation can be explained by a simple concept 
of ``order-parameter switching'', which originates from the 
general nature of the viscoelastic model.  
In Sec. IX, we discuss the universal nature of spongelike morphology 
characteristic of a dynamically asymmetric mixture containing a fluid as its 
component. In Sec. X, we conclude our work. 
In Appendix, we briefly mention the applications of viscoelastic 
phase separation to material science.

\section{Viscoelastic model based on a two-fluid model}

We review here how a viscoelastic model can be derived 
on the basis of the two-fluid model 
\cite{Doi,DoiOnuki,Milner} 
(see Ref. \cite{DoiOnuki} on the details of the theoretical method). 
The model has originally be derived to describe the phase separation 
in polymers \cite{DoiOnuki,Milner}, but we believe that the viscoelastic model 
should describe phase separation in any dynamically asymmetric mixture, 
irrespective of the microscopic details of a system \cite{HT1,HT2}.  
Thus, we focus here our special attention on how the most general version 
of the viscoelastic model can be derived. 
Let us consider a two-fluid model of a mixture of components 1 and 2. 
Let $\vec{v}_1(\vec{r},t)$ and $\vec{v}_2(\vec{r},t)$ be the average velocities 
of components 1 and 2, respectively, and $\phi(\vec{r},t)$ 
be the volume fraction of the component 1 at point $\vec{r}$ and time $t$. 
Here we assume the two components have the same density $\rho$ 
for simplicity. 
Then, the conservation law gives 
\begin{eqnarray}
\frac{\partial \phi}{\partial t}=-\vec{\nabla}\cdot(\phi \vec{v}_1)
=\vec{\nabla}\cdot[(1-\phi) \vec{v}_2]
\end{eqnarray}
The volume average velocity $v$ is given by 
\begin{eqnarray}
\vec{v}=\phi\vec{v}_1 +(1-\phi)\vec{v}_2. \label{vave}
\end{eqnarray}
The free energy of the system $F_{mix}$ is given by 
\begin{eqnarray}
F_{mix}=\int d\vec{r} [f(\phi(\vec{r}))+\frac{C}{2}(\nabla \phi(\vec{r}))^2], 
\end{eqnarray}
where $f(\phi)$ is the free energy per unit volume of a mixture 
with the concentration $\phi$ of the component 1.  
The form of $f(\phi)$ depends upon the system; for example, 
it is given by the Flory-Huggins free energy \cite{deGennes} 
for polymer mixtures.  
Its time derivative can be written as 
\begin{eqnarray}
\dot{F}_{mix}&=&\int [\frac{\partial f}{\partial \phi}-C \nabla^2 \phi] 
\dot{\phi} d\vec{r} \nonumber \\
&=&-\int [\frac{\partial f}{\partial \phi}-C \nabla^2 \phi] 
[\vec{\nabla}\cdot (\phi \vec{v}_1)] d\vec{r} \nonumber \\
&=& \int (\vec{\nabla}\cdot \mbox{\boldmath$\Pi$}) \cdot 
\vec{v}_1 d\vec{r},  
\end{eqnarray} 
where $\vec{\nabla} \cdot \mbox{\boldmath$\Pi$}= \phi 
\nabla(\frac{\partial f}{\partial \phi}-C \nabla^2 \phi)$ 
and $\mbox{\boldmath$\Pi$}$ is the osmotic tensor. 
We also assume here that the forces $\vec{F}_i$ acts on the component $i$. 
Thus, the Rayleighian to be minimized is 
\begin{eqnarray}
R=\int d\vec{r} [\frac{1}{2}\rho \frac{\partial}{\partial t} \vec{v}^2+ 
\frac{1}{2} \zeta(\phi)(\vec{v}_1-\vec{v}_2)^2 +
(\vec{\nabla}\cdot \mbox{\boldmath$\Pi$}) \cdot 
\vec{v}_1 \nonumber \\
-p\vec{\nabla}\cdot \vec{v}-\vec{v_1}\cdot \vec{F}_1-
\vec{v}_2 \cdot \vec{F}_2]. 
\end{eqnarray} 
In the above, the term containing the pressure $p$ is added 
to guarantee the incompressibility condition 
\begin{eqnarray}
\vec{\nabla}\cdot \vec{v}=0. \label{incomp}
\end{eqnarray}
The condition that the functional derivatives of the Rayleighian with 
respect to $\vec{v}_1$ and $\vec{v}_2$ be zero gives the following 
equations of motion: 
\begin{eqnarray}
\rho \frac{\partial (\phi \vec{v}_1)}{\partial t}=
-\vec{\nabla}\cdot \mbox{\boldmath$\Pi$}
-\zeta (\vec{v}_1-\vec{v}_2) +\phi \nabla p
+\vec{F}_1, \\
\rho \frac{\partial [(1-\phi) \vec{v}_2]}{\partial t}=
\zeta (\vec{v}_1-\vec{v}_2) +(1-\phi) \nabla p +\vec{F}_2. 
\end{eqnarray}
Thus, the average velocity $\vec{v}$ obeys 
\begin{eqnarray}
\rho \frac{\partial \vec{v}}{\partial t}=
-\vec{\nabla}\cdot \mbox{\boldmath$\Pi$}+\nabla p 
+\vec{F}_1+\vec{F_2}. \label{vdynamics}
\end{eqnarray}
In the quasi-stationary condition, the velocity difference 
between the two components, on the other hand, obeys 
\begin{eqnarray}
\vec{v}_1-\vec{v}_2=-\frac{1}{\zeta}[(1-\phi) 
\vec{\nabla}\cdot \mbox{\boldmath$\Pi$} -(1-\phi)\vec{F}_1+\phi\vec{F}_2]. 
\end{eqnarray}

\section{Coupling of internal modes of material with deformation 
and the division of the resulting stress}
\subsection{Origins of stress}
To obtain explicitly the form of the forces $\vec{F}_i$, 
we need to understand how the stress is partitioned between the two components 
and also to have the microscopic expression of the stress tensor 
of the material. 
The macroscopic total force $\vec{F}$ should be related 
to $\vec{F}_1$ and $\vec{F}_2$ as 
\begin{eqnarray}
\vec{F}=\vec{\nabla}\cdot \mbox{\boldmath$\sigma$}=\vec{F}_1+\vec{F_2}. 
\label{F1F2}
\end{eqnarray}
Here $\mbox{\boldmath$\sigma$}$ is the total stress tensor, which 
is, in general, given by the constitutive equation of material. 
In a linear-response regime, the most general expression 
of $\sigma_{ij}$ is formally written by introducing the 
time dependence of bulk and shear moduli in the theory of elasticity 
\cite{Landau} as 
\begin{eqnarray}
\sigma_{ij}&=&\int^{t}_{-\infty} dt' [G(t-t')\kappa^{ij}_r(t') \nonumber \\
&+&K(t-t') (\vec{\nabla}\cdot \vec{v}_r (t')) \delta_{ij}], \label{sigma}
\end{eqnarray}
where 
\begin{eqnarray}
\kappa_r^{ij}=\frac{\partial v_r^{j}}{\partial x_i} +
\frac{\partial v_r^{i}}{\partial x_j} -\frac{2}{d} 
(\nabla \cdot \vec{v}_r)\delta_{ij}. 
\end{eqnarray}
Here $\vec{v}_r$ is the velocity relevant to the rheological 
deformation and $d$ is the spatial dimensionality. 
$G(t)$ and $K(t)$ are material functions, which we call 
the shear and bulk relaxation modulus, respectively. 
Here it should be noted that $K(t)$ does not contain the bulk 
osmotic modulus, $K_{os}=\phi^2 (\partial^2 f/\partial \phi^2)$. 
We have the relation $\eta= \int_{0}^{\infty} G(t) dt$, 
where $\eta$ is the viscosity of the material. 

{\it The second term of Eq. (\ref{sigma}) is newly introduced 
to incorporate the effect of volume change into 
the stress tensor} \cite{HT3,TA}.  
In a two-component mixture, the mode associated with 
$\vec{\nabla}\cdot \vec{v}_r$ can exist as far as $\vec{v}_r \neq \vec{v}$, 
even if the system is incompressible.  
It should be stressed that its diagonal nature leads to the direct coupling 
with diffusion: note that the effective osmotic pressure is given by 
$\pi^{eff}=(\phi \frac{\partial f}{\partial \phi}-f)-
\int^{t}_{-\infty} dt' K(t-t') \vec{\nabla} \cdot \vec{v}_r(t')$. 
We believe, thus, that this term is 
important even in the case of polymer solution, as described later, although 
this term has so far been ignored (or, more strictly, not treated as an 
important physical factor) in the previous theories 
\cite{Doi,DoiOnuki,Milner,OnukiN}.

\subsection{Estimation of the stress division parameter $\alpha_k$}
Here we consider the physical meaning of $\vec{v}_r$. 
In a linear-response regime, 
the rheological velocity 
$\vec{v}_r$ is generally given by the linear combination of $\vec{v}_1$ and 
$\vec{v}_2$ \cite{DoiOnuki,OnukiN}:
\begin{eqnarray}
\vec{v}_r=\alpha_1 \vec{v}_1 +\alpha_2 \vec{v}_2. \label{vr}
\end{eqnarray}
Then, the next problem is how the stress is partitioned 
between the two components. 
Since $\vec{F} \cdot \vec{v}_r$ 
should be equal to $\vec{F}_1\cdot \vec{v}_1+\vec{F}_2\cdot \vec{v}_2$ in the 
Rayleighian, we have the following stress division: 
\begin{eqnarray}
\vec{F}_1=\alpha_1 \vec{F}, \quad \vec{F}_2=\alpha_2 \vec{F}. \label{stressdiv}
\end{eqnarray}
Here $\alpha_1+\alpha_2=1$ from Eq.~(\ref{F1F2}). 

\subsection{Direct Estimation of $\vec{F}_k$}
Here we consider the meaning of the forces from a different viewpoint. 
The forces acting on the component $k$ are (i) the friction between 
the component $k$ and the other component due to their relative motion and 
(ii) the rheological coupling between the component $k$ and the surrounding 
rheological environment including the component $k$ itself. 
This can be easily understood by considering gel that is composed of 
polymer network and solvent, as an example: 
The motion of polymer is affected by the two forces, namely, the friction 
force against solvent and the elastic force due to the network deformation. 
Thus, 
{\it it is natural to think that $\vec{F}_k$ {\rm ($k=1,2$)} 
corresponds to the force of type (ii), namely, the force 
acting on the component $k$ by the motion of the component $k$ 
{\rm ($\vec{v}_k$)} itself}, and not by that of the other component. 
Thus, we assume that $\vec{F}_k$ is linear in $\vec{v}_k$: 
\begin{eqnarray}
\vec{F}_{k}&=&\vec{\nabla}\cdot \mbox{\boldmath$\sigma$}^{(k)}, \\
\sigma^{(k)}_{ij}&=&
\int^{t}_{-\infty} dt' [G^{(k)}(t-t')\kappa_k^{ij}(t') \nonumber \\
&\quad&+K^{(k)}(t-t') (\vec{\nabla}\cdot \vec{v}_k(t')) \delta_{ij}], 
\label{sigmaeach}
\end{eqnarray} 
where
\begin{eqnarray}
\kappa_k^{ij}=\frac{\partial v_k^{j}}{\partial x_i} +
\frac{\partial v_k^{i}}{\partial x_j} -\frac{2}{d} 
(\nabla \cdot \vec{v}_k)\delta_{ij}. 
\end{eqnarray}
Here the unknown factors become 
the functional shapes of $G^{(k)}(t)$ and $K^{(k)}(t)$ 
for the motion of the component $k$, instead of $\alpha_k$. 

Using the stress division parameters $\alpha_k$, we obtain the 
following relations between $G$ and $G^{(k)}$ and also 
between $K$ and $K^{(k)}$: 
\begin{eqnarray}
G=\frac{G^{(1)}G^{(2)}}{\alpha_2^2 G^{(1)}+\alpha_1^2 G^{(2)}}, \\
K=\frac{K^{(1)}K^{(2)}}{\alpha_2^2 K^{(1)}+\alpha_1^2 K^{(2)}}.  
\end{eqnarray}
These functions $G^{(k)}$ and $K^{(k)}$ 
express the rheological properties 
of the material responding to the velocity $\vec{v}_k$. 
The fact that $\vec{v}_1$ and $\vec{v}_2$ are coupled with each other 
makes the estimation of the rheological functions difficult. 

\subsection{$\alpha_k$ or $\vec{F}_k$}
The estimation of the stress division of $\vec{F}$ 
into $\vec{F}_1$ and $\vec{F}_2$, 
which is given by the stress division parameter $\alpha_k$,  
formally looks simpler than the direct estimation of the 
above rheological functions, $G^{(k)}$ and $K^{(k)}$; this is true for 
a nearly symmetric stress division, as in the case of 
polymer mixtures: More strictly speaking, when the dynamics 
of both components is governed by the same mechanism, 
$\alpha_k$ is useful to estimate the stress division. 

However, the physical meaning 
of the latter is clearer than the former in a case of 
strongly asymmetric stress division: 
In polymer solutions, for example, the mechanism of 
polymer dynamics is essentially different from 
that of solvent dynamics; and, thus, the rheological functions 
$G^{(k)}$ and $K^{(k)}$ are more useful than $\alpha_k$ (see Sec. VI). 

\section{Basic equations of a viscoelastic model}
Here we summarize the basic equations describing a viscoelastic model: 
\begin{eqnarray}
\frac{\partial \phi}{\partial t}&=&-\vec{\nabla}\cdot(\phi \vec{v}) 
-\vec{\nabla}\cdot [\phi(1-\phi) (\vec{v}_1-\vec{v}_2)], \label{k1}
\\
\vec{v}_1-\vec{v}_2&=&-\frac{1-\phi}{\zeta}[
\vec{\nabla}\cdot \mbox{\boldmath$\Pi$} -\vec{\nabla}\cdot  
\mbox{\boldmath$\sigma$}^{(1)}+\frac{\phi}{1-\phi}\vec{\nabla}\cdot  
\mbox{\boldmath$\sigma$}^{(2)}], \label{k2} \\ 
\rho \frac{\partial v}{\partial t} &\cong& 
-\vec{\nabla}\cdot \mbox{\boldmath$\Pi$}+\nabla p 
+\vec{\nabla}\cdot \mbox{\boldmath$\sigma$}^{(1)}+
\vec{\nabla}\cdot \mbox{\boldmath$\sigma$}^{(2)}.  \label{k3}
\end{eqnarray}
We also need Eq. (\ref{vave}), Eq. (\ref{incomp}), 
Eq. (\ref{sigmaeach}), and the information on the stress division. 
Here it should be noted that we need the phenomenological or microscopic 
theories describing the forms of $G(t)$ and $K(t)$ or those of 
$G^{(k)}$ and $K^{(k)}$. 
Since we derive the above basic equations, relying only upon 
a two-fluid model, {\it they should be independent 
of types of material and quite general}. 

\section{Asymmetric stress division}
Here we first focus on some specific problems to have 
more deep insight on the origins of stress and how the resulting stress 
is partitioned between the two components of a mixture. 
Then, we consider some general features of asymmetric stress division.   

\subsection{Examples of asymmetric stress division}

\subsubsection{Polymer solutions: cases of good and $\theta$ solvents} 
In this case, the stress division has also already been given 
in literature \cite{Doi,DoiOnuki,Milner,OnukiN}: 
$\alpha_1 \cong 1$ and $\alpha_2 \cong 0$, provided 
that the component 1 is a polymer and 2 is a solvent. 
However, this division itself is based indirectly on the estimation 
of $G^{(k)}$ since the stress division parameters cannot be 
determined precisely. 
It should be stressed that there is no direct way to determine $\alpha_1$ and 
$\alpha_2$ from the first principle. Thus, there is no firm basis 
for $\vec{v}_r=\vec{v}_1$ in polymer solution, 
as pointed out by Doi and Onuki \cite{DoiOnuki}, although it looks 
natural physically. In this case, we believe that Eq.~(\ref{sigmaeach}) 
is more useful and physically easier to understand than Eq.~(\ref{sigma}), 
as described in Sec. III. D.

The topological entanglement can be felt only by polymers, 
and never by solvent molecules. 
Then $G^{(1)}(t)$ is approximately 
given by the existing polymer-solution theory \cite{Doibook} 
[$G^{(1)}(t) \cong G(t)$]
and $K^{(1)}(t) \sim 0$ \cite{Doi,DoiOnuki,Milner}, 
{\it if the solvent is not poor}. 
On the other hand, $G^{(2)}(t)$ is an extremely fast decay function 
and, thus, the solvent viscosity $\eta_2$ is obtained as 
$\eta_2=\int^{\infty}_{0}dt G^{(2)}(t)$. 
Since the solvent has no large internal degrees of freedom, 
we can safely assume that $K^{(2)}=0$. 

\subsubsection{Polymer solutions: a case of poor solvent}
It should be stressed that phase separation of polymer solution always occurs 
in a poor solvent; and, thus, the case of poor solvent is extremely important 
when we consider critical phenomena and phase-separation phenomena. 
Unfortunately, however, there do not exist any established theories 
that describe quantitatively the polymer dynamics in a poor solvent. 
In a poor solvent, we need to consider the attractive interactions 
between polymer chains seriously. Thus, the most natural model 
is a transient gel model \cite{HT3,TA} in which the interpolymer 
attractive interaction 
produces the temporal contact point (crosslinking) between polymer chains. 
If we assume that the life time of the temporal contact between chains is 
$\tau_x$, we expect that the bulk relaxational 
modulus $K^{(1)}(t)$ has a relaxation time of the order of $\tau_x$. 
$\tau_x$ likely obeys the Arrhenius-type law: 
$\tau_x=\tau_x^0 \exp(E/k_BT)$ ($k_B$: the Bolzmann constant). 
Here the bonding energy $E$ is likely proportional 
to $T_\theta-T$ ($T_\theta$: the $\theta$ temperature) near 
$T_\theta$, and also dependent upon the distance and orientation 
of the relevant segments. 
Even in polymer solutions, thus, we expect that $K^{(1)}(t)$ plays 
an important role in viscoelastic phase separation in contrast to 
the previous theories \cite{Doi,DoiOnuki,Milner,OnukiN}: 
{\it The transient network of topological origin itself (entanglement 
effects) might not lead to the bulk relaxation mode, while the transient 
network formed by attractive interactions does clearly 
makes the bulk relaxation mode active. } 
We believe that it is this mode that is primarily responsible for 
the volume shrinking behavior of a more viscoelastic phase 
observed in our experiments \cite{HT1,HT2}. 
The characteristic decay time of $G^{(1)}(t)$ is likely 
longer than $\tau_x$ since the reptation-like motion 
is additionally required for the polymer motion 
under the temporal network formed by the attractive interaction. 
We need further theoretical studies to have quantitative 
expressions of $G(t)$ and $K(t)$ in a poor solvent. 

Relating to the above transient gel model, 
we speculate that a polymer solution behaves as 
physical gel {\it universally} at least at a high polymer 
concentration under a strongly poor-solvent condition. 
The existence of special junction points is not a prerequisite 
to the formation of such a transient gel. 
The transient pairing of any parts of two chains can 
be regarded by a temporal crosslinking.  
Any pair of segments of polymer chains can form a temporal crosslinking 
point, irrespective of inter or intrachain. The probability 
of its formation is determined by the balance between the 
intersegment attractive interaction depending upon the geometrical 
configuration of chains and the thermal energy. 
Thus, the most probable candidate of the contact point 
is an entanglement point. 
The sol-gel transition can be given simply by 
the criterion that the transient network formed by interpolymer 
attractive interaction is percolated at any moment. 
A polymer chain having at least two contact points with other different 
polymers plays a role as a junction point of usual physical gel. 
If we assume simply that the topological entanglement point 
is the only candidate for a temporal crosslinking point, 
the criterion for physical gelation is given by 
\begin{eqnarray}
\frac{N}{N_e}\exp(E/k_BT) \geq 2, 
\end{eqnarray}
where $N$ is the degree of polymerization of 
polymer and $N_e$ is the degree of polymerization between entanglement points. 
Further quantitative studies along the above line are 
highly desirable.

\subsubsection{Polymer mixtures}
In the case of a mixture of polymers 1 and 2, whose 
degrees of polymerization are $N_1$ and $N_2$, respectively, 
the stress produced by the motion of polymer 1 
can be different from that by the motion of polymer 2. 
Intuitively, the motion of a longer chain causes stronger stress 
than that of a shorter chain does. 
Following the Brochard theory on mutual polymer diffusion \cite{Bro}, 
which is based on the reptation 
theory that mainly deals with the effect of topological 
constraints (tube) on entangled polymer chains,  
Doi and Onuki \cite{DoiOnuki} have explained how the stress 
should be divided by the two 
polymers with different lengths. According to them, 
\begin{eqnarray}
\vec{v}_r&=&\vec{v}_T=\alpha_1 \vec{v}_1+\alpha_2 \vec{v}_2, \\
\alpha_k&=&\frac{\zeta_k}{\zeta_1+\zeta_2}= 
\frac{\phi N_k}{\phi N_1+(1-\phi) N_2},  
\end{eqnarray} 
where $\vec{v}_T$ is the tube velocity. 
Here $\zeta_k$ is the friction of the component $k$ with 
the tube surrounding it. 
$\zeta_k$ is given by $\zeta_k=\phi_k (N_k \zeta_0/N_e)$ 
\cite{DoiOnuki}, 
where $\phi_k$ is the volume fraction of the component $k$, 
$\zeta_0$ is the microscopic friction constant, and $N_e$ 
is the average degree of polymerization between the entanglement 
points. 
The resulting stress division is given by 
$\vec{F}_k=\alpha_k \vec{\nabla}\cdot \mbox{\boldmath$\sigma$}$ \cite{DoiOnuki}. Here it should be stressed that 
{\it $\zeta_k$ is the very essential quantity in the sense that 
it represents the coupling strength between the component $k$ 
(the volume fraction of $\phi_k$) and the surrounding rheological 
environment.} 

Near and below a critical point $T_c$, however, we also have to consider 
the role of the attractive interactions between the same kind 
of polymers, which increases the rheological coupling, namely, 
$\zeta_k$. For example, this leads to the slower diffusion constant 
than that predicted by a reptation theory which concerns only 
topological effects and neglects energetic interactions between 
polymers. The inclusion of energetic interactions is a prerequisite 
to the more precise description of polymer dynamics. 
However, the following fact should be mentioned: 
High-molecular weight polymer mixtures often mix at a lower 
temperature, and demix at a higher temperature. In such a case, 
the energetic interactions likely play more important roles 
in polymer dynamics in the one-phase region rather 
than in the demixing region.

\subsubsection{A mixture of components having very different 
glass-transition temperatures ($T_g$)}
In this case, we also expect an asymmetric stress division 
since the two kinds of the component molecules are expected to 
feel very differently the rheological environment as in the case 
of polymer mixtures, even if the mean-field rheological environment 
surrounding them is the same. 
We have actually observed viscoelastic phase separation in a mixture 
of polymers having very different $T_g$, whose behavior is essentially 
the same as that of polymer solutions \cite{HT2}. 
It is easy to imagine that a high $T_g$ component has less 
friction with the local rheological environment 
than a low $T_g$ component. 
Recent theoretical studies on supercooled binary liquids \cite{Bosse} 
based on the mode-coupling approximation support 
such a picture. 
If we introduce formally the coupling strength 
$\zeta_k$ for the component $k$ that is proportional to 
$\phi_k$, the stress division 
can be expressed by the same relation as 
in the above case of polymer mixtures. 
The mean-field rheological environment in this problem 
of glass transition is the so-called ``cage'' \cite{Richert}. 
{\it The concept of ``cage'' in glass transition is quite similar to the 
concept of ``tube'' in polymer mixtures.} 
The escape time of a molecule from ``cage'' or ``tube'' gives 
the relaxation time of $G(t)$ in both cases. 
Unfortunately, we do not have the reliable theoretical 
basis even for a simple liquid-glass transition; and, thus, 
it is difficult to have specific quantitative expressions 
for $\zeta_k$ at present. 
Phenomenologically, however, it is known that 
$G(t)=G_0 \exp[-(t/\tau)^\beta]$ and 
$\tau \sim \tau_0 \exp(B/(T-T_0))$, where $\beta$ is the stretching 
parameter ($0 \le \beta \le 1$) and $T_0$ is the so-called 
Vogel-Fulcher temperature. 

It should be stressed again that we have to take into account 
the effects of attractive interactions between 
the same species on their dynamics below $T_c$ \cite{HT4}. 

\subsubsection{Colloidal suspensions}
It is well known that the addition of enough non-absorbing polymer 
to an otherwise stable colloidal suspension can induce phase separation 
{\it via} the depletion mechanism. 
Colloidal suspensions form a transient gel state in the initial 
stage of phase separation \cite{Poon} 
in much the same way as polymer solutions do. 
We believe that the essential 
features of colloid phase separation can also be well described 
by our viscoelastic model. The dynamic asymmetry in colloidal suspensions 
simply comes from the size difference between colloids and 
solvent molecules. 

\subsection{Physical origin of asymmetric stress division}
Here we consider a problem of what is the most basic physical factor 
that is responsible for asymmetric stress division, 
on the basis of intermolecular or interparticle interactions. 
The network of attractive interaction is universally formed when a 
mixture is quenched into its metastable or unstable state since  
there exist attractive interactions between the same components. 
In dynamically symmetric mixtures, the interaction network 
always relaxes in its equilibrium state much faster than the 
phase-separation process. 
In dynamically asymmetric mixtures, however, 
the relaxation time of the interaction network is different 
between the two components because of the mobility difference.  
This consideration based on microscopic interactions leads to 
the conclusion that {\it the dynamic asymmetry between the 
components of a mixture is the essential origin of 
asymmetric stress division}. 
Thus, the phase-separation behavior of any dynamically asymmetric 
mixtures including the above 1-5 should be essentially the same 
and described by Eq. (\ref{k1})-(\ref{k3}). 

\subsection{A general rule of stress division}
On the basis of the above examples, we discuss 
a general rule of the stress division in viscoelastic matter. 
Here we do not consider elastic matter where the elastic coupling 
also plays an important role in addition to the friction.
In the preceding discussion, 
we obtain the general relation given by [Eq.~(\ref{vr})], 
$\vec{v}_r=\alpha_1 \vec{v}_1+\alpha_2 \vec{v}_2$, with 
$\alpha_1+\alpha_2=1$. 
For the relative motion of the component $k$ having the velocity 
of $v_k$ to the mean-field rheological environment having 
the velocity of $\vec{v}_r$, the friction force 
is given by $\zeta_k(\vec{v}_r-\vec{v}_k)$, 
where $\zeta_k$ is the average friction 
of the component $k$ and the mean-field rheological environment 
at point $\vec{r}$, where the volume fraction of $k$ component 
is $\phi_k(\vec{r})$. 
Here $\zeta_k=\phi_k \zeta_k^{m}$ and 
$\zeta_k^{m}$ is proportional to the friction between an individual 
molecule of the component $k$ and the mean-field rheological environment, 
which we call the generalized friction parameter. 
Because of the physical definition of the mean-field 
rheological environment, the two friction forces 
should be balanced. 
This fact guarantees that the rheological properties 
can be described only by $\vec{v}_r$ as in Eq.~(\ref{sigma}). 
Thus, we have the following relation, in general: 
\begin{eqnarray}
\zeta_1(\vec{v}_r-\vec{v}_1)+\zeta_2(\vec{v}_r-\vec{v}_2)=0. \label{balance}
\end{eqnarray} 
From Eqs.~(\ref{vr}) and (\ref{balance}), we obtain the 
general expression of the stress division parameter $\alpha_k$:  
\begin{eqnarray}
\alpha_k= \frac{\phi_k\zeta_k^{m}}{\phi_1\zeta_1^{m}+\phi_2\zeta_2^{m}}.  
\label{alphak}
\end{eqnarray}
The above relation is consistent with a simple 
physical picture that the friction is only the origin of the coupling 
between the motion of the component molecules and the rheological medium. 
The above relation is a straightforward extension of the stress division 
in polymer mixtures \cite{DoiOnuki}, where $\vec{v}_r$ is the tube velocity 
$\vec{v}_T$. 
We expect that this relation holds, irrespective of the 
microscopic details of rheological models, and, thus, 
we can apply it to a mixture of any material, the motion of both of whose 
components is described by a common mechanism. 
However, it should be stressed that {\it this relation is not useful 
for mixtures whose components have essentially different 
mechanisms of molecular motion as in the case of polymer solutions}: 
In polymer solutions, for example, the motion of polymers is essentially 
different from that of solvent molecules in the mechanism. 
In the standard rheological theory of polymer solutions, 
$G(t)$ itself is estimated as the sum of the polymer contribution and the 
solvent contribution. Thus, the stress division cannot be simply described 
by Eq.~(\ref{alphak}). In such cases, Eq.~(\ref{sigmaeach}) is more useful, 
as mentioned in Sec. III. D.

\section{Generality of a viscoelastic model}
Next we briefly discuss the generality of the above viscoelastic model 
described by Eqs.~(\ref{k1}), (\ref{k2}), and (\ref{k3}) \cite{HT3}. 
This model including the bulk volume relaxation mode 
is quite a general model, as shown below. 
We describe below how the viscoelastic model reduces 
to various models under some assumptions. 

\subsection{Elastic solid model}

If we assume that $G(t)=\mu(\phi)$ ($\mu$: shear modulus) and $K(t)=K_b(\phi)$ 
($K_b$: bulk modulus) and $\vec{v}=0$, 
this model reduces to the model of elastic solid model \cite{ON}. 
Since the time integration of the velocity 
becomes the deformation $u$, the stress is given by 
\begin{eqnarray}
\sigma_{ij}&=&\mu(\phi)[\frac{\partial u_{j}}{\partial x_i} +
\frac{\partial u_{i}}{\partial x_j} -\frac{2}{d} 
(\vec{\nabla} \cdot \vec{u})\delta_{ij}] \nonumber \\
&+&K_b(\phi) (\vec{\nabla}\cdot \vec{u}) \delta_{ij}. 
\end{eqnarray} 
Thus, the basic kinetic equation is given by 
\begin{eqnarray}
\frac{\partial \phi}{\partial t}=\vec{\nabla}\cdot \frac{\phi (1-\phi)^2}
{\zeta}[\vec{\nabla}\cdot \mbox{\boldmath$\Pi$}&-&\vec{\nabla}\cdot  
\mbox{\boldmath$\sigma$}^{(1)} \nonumber \\
&+&\frac{\phi}{1-\phi} \vec{\nabla}\cdot  \mbox{\boldmath$\sigma$}^{(2)}]. 
\label{ke1} 
\end{eqnarray}
In this case, the softer phase form a networklike phase 
because the deformation of the softer phase costs less energy than that of 
the harder phase \cite{ON}. 
It should be stressed that the force balance condition 
plays no roles in determining the morphology. 
This fact causes the striking difference in morphology between an elastic solid 
model and an elastic gel or asymmetric viscoelastic model \cite{HT1}: 
In the former the softer phase forms the networklike structure, while 
in the latter the harder phase does.

\subsection{Solid model}
If we assume the dynamic symmetry (no dependence of $\mu$ and $K_b$ 
on $\phi$) further, it reduces the solid model 
(model B \cite{HH}). This is because we have the symmetric stress 
division as $(1-\phi)\vec{F}_1=\phi\vec{F}_2$. 
Here it should be noted that the condition $\mu=K_b=0$ 
is unnecessary and only the symmetry in elastic properties between the 
two components is required.
In this case, the basic equation becomes the simplest diffusion 
equation: 
\begin{eqnarray}
\frac{\partial \phi}{\partial t}=\vec{\nabla}\cdot \frac{\phi (1-\phi)^2}
{\zeta}[\vec{\nabla}\cdot \mbox{\boldmath$\Pi$}] 
\label{ks1} 
\end{eqnarray}

\subsection{Symmetric Viscoelastic model}
If we assume only the dynamic symmetry between 
the two components of a mixture, 
it reduces to a new "symmetric viscoelastic model".  
In this case, we have a trivial stress division: 
$\vec{F}_1=\phi \vec{\nabla}\cdot \mbox{\boldmath$\sigma$}$ and 
$\vec{F}_2=(1-\phi) \vec{\nabla}\cdot \mbox{\boldmath$\sigma$}$. 
Namely, $\alpha_1=\phi$ and $\alpha_2=1-\phi$. 
The rheological functions $G^{(k)}$ can be estimated as 
$G^{(1)}(t)= \phi G(t)$ and $G^{(2)}(t)= (1-\phi) G(t)$. 
In this particular case, $\vec{v}_r=\vec{v}$; and, 
accordingly, there should be no contribution of the bulk 
relaxation modulus under the incompressibility condition 
($\vec{\nabla} \cdot \vec{v}=0$). 
The basic kinetic equations are given by 
\begin{eqnarray}
\frac{\partial \phi}{\partial t}&=&-\vec{\nabla}\cdot(\phi \vec{v}) 
+\vec{\nabla}\cdot \frac{\phi (1-\phi)^2}{\zeta} \vec{\nabla}\cdot 
\mbox{\boldmath$\Pi$}, 
\\
\rho \frac{\partial v}{\partial t} &\cong& 
-\vec{\nabla}\cdot \mbox{\boldmath$\Pi$}+\nabla p 
+\vec{\nabla}\cdot \mbox{\boldmath$\sigma$}. \label{fb}
\end{eqnarray}
Since $\vec{v}_r=\vec{v}$, the gross variables describing the dynamics 
are only $\phi$ and $\vec{v}$. 
Here it should be stressed that the rheological function 
$G(t)$ does not depend 
upon the location $\vec{r}$ because of the dynamic symmetry. 
Using the relation $\vec{\nabla} \cdot \vec{v}=0$, thus, 
\begin{eqnarray}
\vec{F}=\nabla \cdot \mbox{\boldmath$\sigma$}=\int^{t}_{-\infty} 
dt' G(t-t') \nabla^2 \vec{v}(t').  
\end{eqnarray}
We have also the relation 
\begin{eqnarray}
\vec{v}_1-\vec{v}_2&=&-\frac{1-\phi}{\zeta}[\vec{\nabla}\cdot 
\mbox{\boldmath$\Pi$}], \label{v1v2} 
\end{eqnarray}
although it is unnecessary for solving the problem. 

This model describes the dynamics of dynamically symmetric 
polymer mixtures. It should be stressed that this model is different 
from a fluid model (model H) described below. Thus, there remains 
a possibility that there is a new polymer effect relating to 
this model: For example, we expect an unusual feature in the initial stage of 
phase separation where the deformation rate is large. 
The polymer effect corresponding to this was first pointed out by 
de Gennes \cite{deGennes1}, and has very recently been studied in detail 
by Kumaran and Frederickson \cite{Frederickson}.

\subsection{Fluid Model}
If we assume that the deformation is much slower than 
the internal rheological time of the material for 
the above model, we further have the relation 
$\vec{\nabla}\cdot \mbox{\boldmath$\sigma$}=\eta \nabla^2 \vec{v}$. 
Thus, the model reduces to the fluid model (model H \cite{HH}):
\begin{eqnarray}
\frac{\partial \phi}{\partial t}&=&-\vec{\nabla}\cdot(\phi \vec{v}) 
+\vec{\nabla}\cdot \frac{\phi (1-\phi)^2}{\zeta} \vec{\nabla}\cdot 
\mbox{\boldmath$\Pi$}, 
\\
\rho \frac{\partial v}{\partial t} &\cong& 
-\vec{\nabla}\cdot \mbox{\boldmath$\Pi$}+\nabla p 
+\eta \nabla^2 \vec{v}. 
\end{eqnarray}

\subsection{Elastic gel model}
If we assume only $G=\mu(\phi)$ and $K=K_b(\phi)$, 
it reduces to the elastic gel model \cite{Sekimoto,Onukigel} 
that describes phase separation 
in elastic gel. 
The basic equations are essentially the same as those 
of viscoelastic phase separation in polymer solution 
[Eqs. (\ref{k1}), (\ref{k2}), 
and (\ref{k3})], except that the stress tensor is given by 
\begin{eqnarray}
\sigma^{(1)}_{ij}&=&\mu(\phi)[\frac{\partial u_1^{j}}{\partial x_i} +
\frac{\partial u_1^{i}}{\partial x_j} -\frac{2}{d} 
(\vec{\nabla} \cdot \vec{u}_1)\delta_{ij}] \nonumber \\
&+&K_b(\phi) 
(\vec{\nabla}\cdot \vec{u_1}) \delta_{ij}, \nonumber \\
\sigma^{(2)}_{ij}&=& \eta_2 [\frac{\partial v_2^{j}}{\partial x_i} +
\frac{\partial v_2^{i}}{\partial x_j} -\frac{2}{d} 
(\vec{\nabla} \cdot \vec{v}_2)\delta_{ij}]. \nonumber 
\end{eqnarray} 
In this case, when the elastic energy overcomes the mixing free energy, 
phase separation stops and the coarsening of domains is pinned. 

\subsection{Generality and intrinsic nonuniversality}

Since any phase separation in all isotropic condensed matter 
can be classified into solid, elastic solid, elastic gel, 
symmetric and asymmetric 
viscoelastic, and fluid models, the above viscoelastic model (see Sec. IV) 
including both shear and bulk relaxation stresses should be a universal 
model describing phase separation and critical phenomena 
in isotropic matter without any exception. 

However, this model is not universal in the usual sense 
of critical phenomena since 
it requires some microscopic theories describing the 
rheological properties of the matter.  
In the extreme limit of the strong dynamic asymmetry, 
the elementary slow dynamics (internal mode) of material affects 
the critical fluctuation even near the critical point, 
in contrast to the concept of the dynamic universality \cite{HH}. 
Thus, there is a possibility that we cannot experimentally 
approach to a critical regime where  
the order parameter dynamics is only the slow mode 
in the system \cite{HT01,HT1}. 

The viscoelastic effects can be parameterized by the so-called 
viscoelastic length $\xi_{ve} \sim (D_\xi \tau_t)^{1/2}$ 
originally introduced by Brochard and de Gennes 
\cite{BD}, 
where $\tau_t$ is the characteristic time of rheological 
relaxation and $D_\xi$ is the diffusion constant. 
For the length scale longer than $\xi_{ve}$, concentration fluctuations 
decay by diffusion, while in the length scale shorter 
than viscoelastic effects dominate \cite{BD,DoiOnuki,Milner}. 
This is a simple mapping of the dynamic crossover 
that in the time scale longer than $\tau_t$ diffusion dominates 
concentration fluctuations while in the time scale shorter 
than $\tau_t$ viscoelastic effects dominate. 
The critical regime is, thus, described by the condition 
$\xi \gg \xi_{ve}$, where $\xi$ is the correlation length 
of concentration fluctuations. 
This condition can also be written as $\tau_\xi \gg \tau_t$, 
where $\tau_\xi$ is the characteristic time of the 
critical concentration fluctuations. 
Thus, we need to consider whether we can easily approach to 
the critical regime that is defined by the above criterion, 
in a viscoelastic system. 
The more detailed consideration on this problem of 
{\it intrinsic nonuniversality}  
will be described elsewhere. 

\section{Viscoelastic suppression of local concentration fluctuations 
in polymer solutions}

\subsection{Fluctuation suppression due to bulk relaxation modulus}
According to the continuity equation, Eq. (1), we have the following relation, 
\begin{eqnarray}
\frac{\partial \phi}{\partial t}= -\nabla \phi \cdot \vec{v}_1 -
\phi \vec{\nabla}\cdot \vec{v}_1. \label{df} 
\end{eqnarray}
Here we assume that the component 1 is a polymer and the component 
2 is a solvent.  
In the above equation, the first term of the right-hand side simply describes 
the translational transport of polymers to a point $\vec{r}$ 
by the locally uniform velocity field $\vec{v}_1$, while 
the second one describes the polymer 
diffusion toward or from a point $\vec{r}$. 
Thus, the first term is associated only with the change in the spatial pattern 
of the concentration distribution, while the second term is responsible 
for the change in the concentration distribution itself.  
In the initial stage of phase separation, the major process 
is the diffusion process leading to 
the change in the concentration distribution itself and 
there are few changes in the spatial pattern of the concentration distribution. 
Neglecting the first term in Eq. (\ref{df}), thus, we have the relation 
\begin{eqnarray}
\frac{\partial \phi}{\partial t}/\phi \cong -\vec{\nabla}\cdot \vec{v}_1. 
\end{eqnarray}
The left-hand side of the above equation is inversely 
proportional to the characteristic time of the concentration change, 
$\tau_\phi$. 
On the other hand, the bulk relaxation modulus $K(t)$ that 
is directly coupled with $\vec{\nabla}\cdot \vec{v}_1$ 
has a characteristic 
decay time of $\tau_x$, which is related to the characteristic time of the 
transient crosslinking between polymer chains in a poor-solvent condition. 
If $\tau_x \gg \tau_\phi$, the rapid growth of concentration fluctuations 
characteristic of spinodal decomposition is suppressed severely 
and may even be prohibited. 
If $\tau_x \ll \tau_\phi$, on the other hand, there are few elastic effects 
and the concentration fluctuations can grow as in usual spinodal decomposition. 

\subsection{Spinodal decomposition {\it vs.} nucleation and growth} 
If the above mechanism of the suppression of concentration fluctuations 
works efficiently, spinodal decomposition cannot proceed further 
after the crossover between $\tau_x$ and $\tau_\phi$. Thus, 
the type of phase separation switches from spinodal decomposition 
to nucleation and growth. 
More strictly speaking, the concentration fluctuations 
grow only locally in the solvent-rich region where $\tau_x$ is short 
and the diffusion is easy to take place. 
Such behavior is actually observed in our experiments \cite{HT1}. 
The crossover from an initial fluid state to a transient gel state 
likely takes place almost immediately after the temperature quench, since 
the long-range diffusion or motion of molecules are not required for 
the formation of a transient network (see also Sec. VIII. C.). 
After the formation of a transient gel state, the 
``mechanical instability'' that is a universal feature of the ``soft'' network 
of attractive interactions leads to the nucleation-growth-like behavior.   
In the following, we discuss the concept of "order-parameter switching" 
resulting from the crossover between the characteristic phase-separation time 
and the internal rheological time.

\section{Order-parameter switching}

\subsection{Order-parameter switching between composition and 
deformation tensor}
Here we consider the dynamic process of viscoelastic phase separation 
on the basis of the viscoelastic relaxation phenomena 
described by Eq.~(\ref{sigma}). 
The quantitative feature of the dynamics can be understood on the basis 
of a concept of ``order-parameter switching'' \cite{HT3}. 
Phase separation is usually driven by a thermodynamic force 
and the resulting ordering process can be described by the temporal 
evolution of the relevant order parameter associated with 
the thermodynamic driving force. 
The primary order parameter describing phase separation 
of a binary mixture is a composition difference 
between the two phases. 
Besides exceptional cases where phase separation and other 
ordering processes such as superfluidization, gelation, 
liquid-crystallization, and crystallization, simultaneously 
proceed \cite{Gunton} [in other words, 
there are more than two kinds of thermodynamic forces 
(order parameters)], 
a phase-separation process is usually characterized by 
a single order parameter.  
In the viscoelastic model, on the other hand, 
the phase-separation mode can be switched 
between ``fluid mode'' and ``elastic gel mode''.  This switching is likely 
caused by the change in the coupling between stress fields and 
velocity fields, which is described by Eq.~(\ref{sigma}): 
Equation~(\ref{sigma}) tells us that 
these two ultimate cases, namely, (i) fluid model 
($\kappa_{ij}^{p}, \vec{\nabla}\cdot \vec{v}_r \sim const$) and 
(ii) elastic gel model ($G(t), K(t)\sim const$), 
correspond to $\tau_{ts} \gg \tau_d$ 
and $\tau_{ts} \ll \tau_d$, respectively. 
Here $\tau_d$ is the characteristic time of deformation, and $\tau_{ts}$ 
is the characteristic rheological time of the slower phase. 
There can be two types of $\tau_{ts}$: One is associated with 
the characteristic decay time of $G(t)$, while the other 
with that of $K(t)$. We think that the former is generally longer 
than the latter in polymer solutions, as mentioned in Sec. V. A. 2.
As described in the preceding section, we believe that the latter 
plays an important role in the suppression of 
concentration fluctuations in the initial stage. 

For $\tau_d \gg \tau_{ts}$ the primary order parameter is the composition 
in usual classical fluids, 
while for $\tau_d \leq \tau_{ts}$ it is the deformation tensor 
($d_{ij}=\partial u_j/\partial x_i+\partial u_i/\partial x_j$) as 
in elastic gels. 
In the elastic regime, the force terms can be included in the Hamiltonian 
as in the case of gel. 
Then the free energy functional is formally written only by 
the deformation tensor $d_{ij}$ as $f(d_{ij})$. 
Thus, we can say that the order-parameter switching 
is a result of the competition between two time scales characterizing 
domain deformation $\tau_d$ and the rheological 
properties of domains $\tau_{ts}$. 
This is a kind of {\it  viscoelastic relaxation} in pattern evolution. 

\subsection{How does the order-parameter switching occur?}
We next consider how  $\tau_{ts}$ and $\tau_d$ change with time during 
phase separation. 
In the initial stage, 
the velocity fields grow as 
$v \sim (k_BT K/3 \eta \xi)\Delta \phi^2$, where $\Delta \phi$ 
is the composition difference between the two phases, and $\xi$ 
is the correlation length, or the interface thickness. 
Since $\Delta \phi$ approaches to $2 \phi_e$ 
($\phi_{e}$ :the equilibrium composition) with time, 
this expression of $v$ reduces to the well-known relation 
$v \sim \gamma/\eta$ ($\gamma$: interface tension) 
in the late stage [note that $\gamma 
\sim k_{\rm B}T K (2\phi_e)^2/3 \xi$].   
Thus, the characteristic deformation time $\tau_d$
changes with time as 
$\tau_d \sim R(t)/v(t) \sim R(t)/\Delta \phi(t)^2$. 
In the initial stage, the domain size does not grow so much with time 
while $\Delta \phi$ rapidly increases with time; 
and, accordingly, $\tau_d$ decreases rapidly. 
On the other hand, $\tau_{ts}$ increases steeply 
with an increase in $\Delta \phi$, reflecting the 
increase in the polymer concentration in a polymer-rich domain. 
Thus, $\tau_{ts}$ becomes comparable to $\tau_d$ in the intermediate stage 
of phase separation. 
Once $\tau_d < \tau_{ts}$, the slower phase cannot follow the deformation 
speed and behaves as an elastic body: The elastic energy dominates 
the coarsening process in the intermediate stage. 
Next we consider the late stage. 
Since $\Delta \phi$ approaches to $2 \phi_e$ and becomes almost constant 
in the late stage, $\tau_d$ ($ \sim R \eta/\gamma$) 
increases with an increase in $R$ while $\tau_{ts}$ becomes almost constant. 
Thus, $\tau_d$ becomes longer than $\tau_{ts}$ again. 
In short, $\tau_d \gg \tau_{ts}$ in the initial stage, 
$\tau_d \leq \tau_{ts}$ in the intermediate stage, and 
$\tau_d \gg \tau_{ts}$ in the late stage again. 
Accordingly, the order parameter switches from the composition to 
the deformation tensor, and then switches back to the composition again. 
This is the first case of ``order-parameter switching'' during an ordering 
process driven by a single thermodynamic driving force, to our knowledge. 

\subsection{Further consideration on $G(t)$ and $K(t)$: 
Kinetics of transient gel formation}
Here we briefly consider viscoelastic functions, $G(t)$ and $K(t)$, 
which are dependent upon material. 
In the above discussion, $\tau_{ts}$ 
is assumed to be a function of $\Delta \phi$. 
However, this picture is not necessary true. 
In polymer solutions and colloidal suspensions, 
for example, the transient gellike structure is likely formed 
very quickly after the quench. 
This is because the diffusion of polymers or colloidal particles 
over a large length scale is not required to form the interaction 
network. The diffusion length scale $l$ is the order of polymer or particle 
size, $a$, near the critical composition ($\phi_c$); 
and, thus, the time required to form network is $\sim a^2/D_a$, where 
$D_a$ is the diffusion constant of a polymer or particle. 
In such a case, $\tau_{ts}$ very rapidly increases 
to the order of $\tau_x$ within a time of $\sim a^2/D_a$ after 
the quench. 
In such a case, the first order parameter switching from composition 
to deformation tensor occurs within a very short period 
($\sim a^2/D_a$) after the quench: The system 
enters into an elastic regime just after the quench. 

The diffusion length scale $l$ increases with a decrease in 
polymer or colloid concentration $\phi$. 
If a percolated network cannot be formed within 
a sufficiently short time, a quasi-homogeneous transient gel state 
is not realized 
due to phase separation, and, thus, a networklike phase-separated 
pattern is not formed; instead, 
a droplet pattern is formed \cite{HT1}. 
This criterion may give the threshold composition between droplet 
and network phase separation. 

Finally, it should be stressed that to describe this network formation 
process and the resulting change in viscoelastic functions, $G(t)$ and $K(t)$, 
we need a microscopic model of each system. 

\subsection{Volume shrinking behavior: 
Absence of self-similar pattern growth in viscoelastic phase 
separation}
Because of the order-parameter switching, 
there is no self-similarity in pattern 
evolution of viscoelastic phase separation. 
In the elastic regime, further, the volume ratio between 
the two phases changes with time \cite{HT1,HT2,HT3}; 
and, thus, there is no proportionality 
between interdomain distance and domain size. 
This behavior even leads to the phase inversion when the more viscoelastic 
phase is a slightly minority phase in equilibrium:
in the initial stage a less viscoelastic phase forms droplets, while 
in the final stage a more viscoelastic phase does. 
This means that there are at least two length scales that change 
differently with time. 
This is also related to the absence of any characteristic length scale 
in elastic deformation. 

Here we make a rough estimate of volume shrinking kinetics. 
The elastic regime should be analogous to the volume shrinking 
of gel. The characteristic shrinking time $\tau_{shrink}$ is likely dependent 
upon the characteristic length scale $L$ as $L^2/D$ 
($D$: gel diffusion constant \cite{TF}). 
It should be noted that only the length scale in this problem 
is the length scale associated with the ``mechanical instability'', 
namely, the characteristic distance between solvent holes, $L_{hole}$. 
Neglecting the time dependence of $L_{hole}$, 
we get a very rough estimation of the volume shrinking time as 
$\tau_{shrink} \sim L_{hole}^2/D$, which can be 
a very long time. The estimation of $L_{hole}$ including 
its time dependence requires 
the stability analysis under the influence of phase separation. 
We need further study on this problem. 

\subsection{Pattern selection: elastic energy {\it vs.} interface energy}
Since the deformation tensor intrinsically has 
geometrical nature, the pattern in the elastic regime 
is essentially different from 
that of usual phase separation in fluid mixtures. 
The domain shape during viscoelastic phase separation 
is determined by which of elastic and interface energy 
is more dominant. 
Roughly, the elastic energy is estimated as $\mu e^2 R^d$ ($e$: strain and 
$d$: spatial dimensions) for a domain of size $R$, 
since it is the bulk energy. 
On the other hand, the interface energy is estimated 
as $\gamma R^{d-1}$. 
For macroscopic domains, thus, the elastic energy 
is much more important than the interface energy 
in the intermediate stage where $\tau_d \leq \tau_{ts}$. 
Accordingly, the domain shape is determined by the elastic force balance 
condition ($\vec{\nabla} \cdot \mbox{\boldmath$\sigma$}^{(n)} \sim0$), 
which leads to networklike or spongelike morphology. 
In the initial and late stages of phase separation where 
$\tau_d \gg \tau_{ts}$, 
on the other hand, the interface energy dominates the domain shape 
since $\mu \sim 0$.

\section{Universality of sponge morphology characteristic 
of a mixture having bulk compression modes}

\subsection{Universality of sponge morphology}
We discuss here the universal nature of a spongelike 
morphology (or the formation of a continuous structure 
by a minority phase) and its physical origin. 
It is known that gel undergoing volume-shrinking 
phase transition forms a bubble-like structure \cite{Sekimoto,Matsuo,Li}. 
The competition between phase separation and gelation or glass transition 
also causes spongelike morphology \cite{Keller,Jackson,Berghmans}. 
The physical origin of the appearance of 
a honeycomb structure in plastic 
foams ({\it e.g.} polystyrene foam and urethane foam) 
is also similar to ours. 
All these processes have a few common features: (i) 
A mixture contains a fluid as a component. (ii) Holes of a less 
viscoelastic fluid phase (gas in plastic foam, water in gel, solvent 
in polymer solution, and so on) are nucleated to minimize 
the elastic energy associated 
with the formation of a heterogeneous structure in an elastic medium. 
(iii) Then, a more viscoelastic phase decreases its volume with time. 
This volume shrinking process is dominated by the transfer (diffusion 
or flow) of a more mobile component under stress fields, 
from a more viscoelastic phase to a less viscoelastic phase. 
The above picture suggests the possibility 
that a spongelike structure is the {\it universal 
morphology} for phase separation in systems in which only one 
component asymmetrically has elasticity stemming from either 
topological connectivity or attractive 
interaction. 

We also point out \cite{HT2} the similarity of these patterns 
in condensed matter to 
the spongelike structure of the universe (the large-scale galaxy distribution) 
\cite{Geller}. 
We speculate that the gravitational 
attractive interaction which is more stronger between heavier matter 
may play a role similar to elastic network in producing the spongelike 
large-scale structure.  
This explanation seems to be consistent with 
a standard picture of the universe evolution 
(a gravitational-instability model) 
that such a heterogeneous structure develops by gravitational 
amplification of density fluctuations. 

This universal appearance of sponge structures in phase separation of 
these systems originates from 
volume phase transition, or more strictly 
elastic phase separation of a {\it dynamically asymmetric mixture 
that is composed of a network-forming component 
and a fluid (such as a liquid and a gas)}. The elastic network 
can be a real one as in gels (permanent network) and 
polymer solutions (transient network) 
or a virtual one due to attractive interactions.  
In the former, the real structure having large internal degrees 
of freedom can store the elastic energy for bulk compression, 
while in the latter the virtual network due to attractive interactions 
can also store the elastic energy. 
In this sense, we can conclude that the existence of both the component 
having bulk (relaxation) modulus and the fluid component 
is a prerequisite for the formation of spongelike structure 
due to the volume shrinking of one phase.

For example, such phenomena are never observed in solid mixtures, 
except for the case that the mobility is strongly dependent 
upon the composition \cite{Jackle}. 
Phase separation of elastic solid mixtures ({\it e.g.} metal alloys) 
does not accompany a drastic volume change of each phase 
if there is no strong composition dependence of mobility. 
This difference causes a marked contrast between elastic phase separation 
in solid mixtures and viscoelastic phase separation; 
in the former, a softer phase always forms a continuous phase 
to minimize the total elastic energy \cite{OnukiN}, 
in contrast to the latter.  

\subsection{Physical origin of volume shrinking}

We briefly discuss the physical meaning of the above criteria 
of the formation of sponge structure, or relative volume shrinking 
of a more viscoelastic phase. 
This is related to the fact that in a two-fluid model 
$\vec{\nabla}\cdot \vec{v}_k$ needs not to be zero 
even under the incompressible condition $\vec{\nabla}\cdot \vec{v}=0$ 
for the average velocity. 
There are three important factors in this problem: 
(i) whether the component $k$ is compressible in a mixture, or not, 
(ii) whether $\vec{\nabla}\cdot \vec{v}_1$ is large enough or not, 
and (iii) whether the change in $\vec{\nabla}\cdot \vec{v}_k$ 
is properly coupled with the stress, or not. 

The condition (i) is usually satisfied since we can change the spatial 
configuration of one component arbitrarily in a two-component mixture, 
in general. 
The condition (ii) is satisfied only for the system containing a fluid as 
its component. 
Finally, the condition (iii) is satisfied only when there exist attractive 
interactions between the components. 
In a simple fluid mixture, for example, $\vec{\nabla}\cdot \vec{v}_k$ 
is not coupled with the elastic stress even if there is a difference 
in viscosity between the two components. 
In relation to this problem, we consider the case of a 
mixture whose components 
have different glass transition temperatures as an example. 
If only one component becomes viscoelastic, the deformation 
of this component that is described by $\kappa_{ij}^{(k)}$ and 
$\vec{\nabla} \cdot \vec{v}_k$ causes the elastic stress. 
Thus, the more viscoelastic phase becomes the matrix phase 
and forms the spongelike structure as experimentally observed \cite{HT2}. 
To prevent usual spinodal decomposition from taking place, the 
bulk modulus should be sufficiently large: The initial growth of 
the concentration fluctuation has to be suppressed mainly by 
$\vec{\nabla}\cdot \vec{v}_k$. For example, this is realized by  
the formation of transient interaction network. 

We believe that the bulk mechanical relaxation modulus $K(t)$ 
plays an essential role in the fluctuation suppression and 
the volume shrinking, while 
the bulk osmotic modulus $K_{os}$ does not play a primary role.  
The interactions can have any 
origins including entropic and energetic ones. 
The long-range nature of the elastic interaction in a more viscoelastic phase 
is directly related to how efficiently the fluctuation is suppressed 
and  the volume of the relevant phase 
can be changed and, thus, to the ability of the formation of 
spongelike structure. 

In relation to the above, it should be mentioned that 
the strong composition dependence of mobility has a similar effect 
\cite{Jackle}, since it slows down the diffusion process in 
the slow-component-rich region selectively. Thus, the similar behavior 
of phase inversion has been observed even in the framework of a solid model 
(model B) \cite{Jackle}. 
As discussed in Sec. V. B, the physical factor 
responsible for the asymmetric stress division in a fluid mixture is 
dynamic asymmetry between the two components of a mixture. 
This strongly indicates that the most essential physical origin of 
volume shrinking behavior and the resulting phase inversion 
is the coexistence of  ``asymmetry in mobility between the 
two components of a mixture'' and ``attractive interactions'', 
irrespective of whether a mixture is solid or fluid.

\subsection{Roles of shear relaxation modulus on the formation 
of a networklike structure}
We have already discussed the roles of bulk relaxation modulus, namely, 
the suppression of the homogeneous growth of concentration fluctuations. 
Here we focus our attention on the roles of shear relaxation modulus. 
An important fact is that the bulk relaxation modulus is closely 
related to the diffusion while the shear relaxation modulus is not: 
The bulk stress gradient $\vec{\nabla}\cdot \mbox{\boldmath$\sigma$}_B$, 
is usually (at least in the initial stage) 
in the same orientation with the osmotic stress gradient, 
$\vec{\nabla}\cdot \mbox{\boldmath$\Pi$}$, since both are 
related to the diagonal part of the deformation velocity, 
$\vec{\nabla}\cdot \vec{v}_r$, as described before. 
On the other hand, the shear stress gradient, 
$\vec{\nabla}\cdot \mbox{\boldmath$\sigma$}_S$, is usually not 
in the same orientation with $\vec{\nabla}\cdot \mbox{\boldmath$\Pi$}$, since 
it is related to the off-diagonal part of the deformation velocity. 
Thus, we think that the shear relaxation modulus 
plays a dominant role in the formation of a networklike 
structure in the intermediate stage of viscoelastic phase separation: 
The overlapping of the stress fields having the spherical symmetry 
around a spherical solvent holes induce the deformation of shear type.  
This initial spherical symmetry of the stress field is characteristic 
of the bulk stress fields coupled with $\vec{\nabla} \cdot \vec{v_r}$. 
The shear deformation causes the shear stress fields through the 
shear relaxation modulus. Thus, the thin part of a more viscoelastic phase 
can support the shear stress and be elongated further.  
In other words, the existence of shear relaxation modulus is responsible for 
the formation of a networklike pattern composed of highly elongated 
thin structures \cite{TA}. 
We believe that without shear relaxation modulus, the networklike 
pattern with threadlike structures can never be formed.  

\subsection{Difference in elastic effects between solid and fluid systems}
Here we discuss the difference in phase-separation morphology 
between elastically asymmetric solid mixtures \cite{ON} and dynamically 
asymmetric fluid mixtures. 

In the diffusion-dominated process, the system approaches to the final 
equilibrium state to reduce the total free energy including the 
elastic energy [see Eq.~(\ref{ke1})]. As a result, the morphology 
that minimizes the elastic energy is selected. 
This is the case of solid mixtures having only elastic asymmetry but no 
dynamic asymmetry. In relation to this, it should be noted that 
the solid mixture having 
dynamic asymmetry behaves entirely differently (see {\it e.g.} Ref. 
\cite{Jackle}).

In the flow-dominated process, on the other hand, 
the force balance condition plays an essential role in pattern selection 
[see Eq.~(\ref{k3})]. 
As a result, the morphology itself is determined by the force 
balance condition. 
The asymmetric stress division leads to the spongelike structure 
where the more viscoelastic phase forms a continuous networklike structure 
to support the stress. 
Further, the two-fluid nature makes the volume change 
of phases possible. 

Thus, we can say that 
dynamic asymmetry is a prerequisite to the phase inversion, 
irrespective of whether material is solid or fluid.

\section{Conclusion}
In summary, we obtain a general model of 
viscoelastic phase separation on the basis of a two-fluid model: 
We demonstrate that the bulk relaxation modulus 
plays an important role in viscoelastic phase separation even 
in polymer solutions. 
Our recent simulations based on the viscoelastic model 
indicates the importance of this bulk mode \cite{TA}. 
Inclusion of this effect makes a viscoelastic model quite general: 
The viscoelastic model can describe phase separation or critical phenomena 
in any isotropic condensed matter without any exception, if there is no 
coupling with additional order parameter. 

Although the viscoelastic model is a quite general model of 
critical phenomena and phase separation, the critical behavior 
of this model may be intrinsically nonuniversal in the sense that 
internal slow modes of material can affect the critical dynamics even near 
the critical point. This problem needs further studies to check whether the 
dynamic universality practically breaks in a mixture having strong dynamic 
asymmetry between its components or not \cite{HT1}. 

As a straightforward extension of the stress division in polymer mixtures 
\cite{DoiOnuki}, we also propose a simple relation describing 
how the stress is divided by the two components, on the basis 
of the idea that the mechanical coupling between a component and the 
mean-field rheological environment is only due to the friction 
between them.  We also point out that the relation is not useful 
for a mixture whose components have 
different mechanisms of the molecular motion. 

We also show that the characteristic features of viscoelastic phase 
separation can be well explained by the concept of "order-parameter 
switching" between composition and deformation tensor. 

We discuss the universal features of spongelike structures observed 
in various material and demonstrate that there is a common 
physical origin that is explained by the framework of our viscoelastic 
model of phase separation. 
It is concluded that the most essential physical origin of 
volume shrinking behavior and the resulting sponge structure 
is the coexistence of ``asymmetry in mobility between the two 
components of a mixture'' and ``attractive interactions'', 
irrespective of whether a mixture is solid or fluid. 
In relation to this, we would like to point out that the phase inversion 
is also observed in a recent simulation of model B including 
the strong composition dependence of the mobility \cite{Jackle}. 
We believe that this is only the way to introduce the dynamic asymmetry 
into a solid model. 
The relation between our model and their model and the underlying 
physics of their similar behavior \cite{TA,Jackle} 
will be discussed in detail elsewhere. 

\section*{Acknowledgments}
The author is grateful for Prof. A. Onuki and Prof. M. Doi 
for valuable discussions. 
He also thanks Prof. J. J\"ackle for discussions on 
the roles of asymmetric mobility and those of bulk relaxation modulus.  
This work was partly supported by a Grant-in-Aid from the 
Ministry of Education, Science, Culture, and Sports, Japan. 

\section*{Appendix A: Comments on applications of viscoelastic 
phase separation in material science}

We briefly discuss the application of 
a spongelike morphology observed in viscoelastic phase separation. 
Although a spongelike structure appear only transiently 
in viscoelastic phase separation, 
this structure can be frozen by suitable methods: (i)simultaneous 
evaporation of a solvent for a polymer solution during phase separation, 
(ii) a further quench of a system below $T_g$, and (iii) 
combination of other processes such as crosslinking reaction. 
We think that spongelike structures reported in 
literature \cite{Keller,Jackson,Berghmans,Song,Widawski} 
are induced primarily by the mechanism described here. 
In relation to this, we point out that some sponge phases 
have periodic structures 
(see {\it e.g.} Ref.\cite{Widawski}), while others do not as in our case. 
This can be explained by the way 
of nucleation of solvent holes: 
only when nucleation is heterogeneously induced 
with a high density in a short period, a periodic sponge structure 
can be formed by the long-range elastic interaction between solvent holes 
(correlated nucleation).  

We also point out that polymerization-induced 
phase separation may lead to the networklike structure 
of a minority phase if there is a certain degree of dynamic 
asymmetry induced by the polymerization of a component. 

In the common sense view of conventional phase separation, 
a minority phase never forms a continuous phase and forms 
only an isolated phase \cite{Gunton}. 
However, our present study indicates the possibility that we can intentionally 
form a spongelike continuous structure of the minority phase 
of a more viscoelastic phase for any 
dynamically asymmetric mixture.

\end{multicols}
\end{document}